\documentclass[5p, times]{elsarticle}

\makeatletter
\def\ps@pprintTitle{%
 \let\@oddhead\@empty
 \let\@evenhead\@empty
 \def\@oddfoot{}%
 \let\@evenfoot\@oddfoot}
\makeatother

\usepackage{hyperref}
\usepackage{subfig}
\usepackage{csquotes}
\usepackage{amsmath}
\usepackage[capitalize]{cleveref}

\usepackage{float}
\usepackage{graphicx}
\usepackage{verbatim}
\usepackage[pagewise]{lineno}
\DeclareUnicodeCharacter{2212}{-}

\begin{document}

\title{Microstructure, grain boundary evolution and anisotropic Fe segregation in (0001) textured Ti thin films}

\author[1]{Vivek Devulapalli \corref{cor1}}
\author[2]{Marcus Hans \fnref{fn1}}
\author[1]{Prithiv T. Sukumar \fnref{fn2}}
\author[2]{Jochen M. Schneider \fnref{fn3}}
\author[1]{Gerhard Dehm \fnref{fn4}}
\author[1]{C.H. Liebscher \fnref{fn5}}
\address[1]{Max-Planck-Institut f\"ur Eisenforschung GmbH, Max-Planck-Stra\ss{}e 1, 40237 D\"usseldorf, Germany}
\address[2]{Materials Chemistry, RWTH Aachen University, Kopernikusstr. 10, D-52074 Aachen, Germany}

\fntext[fn1]{hans@mch.rwth-aachen.de}
\fntext[fn2]{prithiv@mpie.de}
\fntext[fn3]{schneider@mch.rwth-aachen.de}
\fntext[fn4]{dehm@mpie.de}
\fntext[fn5]{liebscher@mpie.de}
\cortext[cor1]{\textbf{Corresponding author: v.devulapalli@mpie.de, Ph: +49 211 6792 434}}

\begin{comment}
\author{Vivek Devulapalli}
\email{v.devulapalli@mpie.de}
\affiliation{Max-Planck-Institut f\"ur Eisenforschung GmbH, Max-Planck-Stra\ss{}e 1, 40237 D\"usseldorf, Germany}

\author{Marcus Hans}
\affiliation{Materials Chemistry, RWTH Aachen University, Kopernikusstr. 10, D-52074 Aachen, Germany}

\author{Prithiv T. Sukumar}
\affiliation{Max-Planck-Institut f\"ur Eisenforschung GmbH, Max-Planck-Stra\ss{}e 1, 40237 D\"usseldorf, Germany}

\author{Jochen M. Schneider}
\affiliation{Materials Chemistry, RWTH Aachen University, Kopernikusstr. 10, D-52074 Aachen, Germany}

\author{Gerhard Dehm}
\affiliation{Max-Planck-Institut f\"ur Eisenforschung GmbH, Max-Planck-Stra\ss{}e 1, 40237 D\"usseldorf, Germany}

\author{Christian H. Liebscher}
\email{liebscher@mpie.de}
\affiliation{Max-Planck-Institut f\"ur Eisenforschung GmbH, Max-Planck-Stra\ss{}e 1, 40237 D\"usseldorf, Germany}
\end{comment}

\begin{keyword}
Titanium; Thin films; Coincident site lattice (CSL); Faceting; Grain boundary segregation; 
\end{keyword}

\begin{abstract}

The structure and chemistry of grain boundaries (GBs) are crucial in determining polycrystalline materials' properties. Faceting and solute segregation to minimize the GB energy is a commonly observed phenomenon. In this paper, a deposition process to obtain pure tilt GBs in titanium (Ti) thin films is presented. By increasing the power density, a transition from polycrystalline film growth to a maze bicrystalline Ti film on SrTiO$_3$ (001) substrate is triggered. All the GBs in the bicrystalline thin film are characterized to be $\Sigma$13 [0001] coincident site lattice (CSL) boundaries. The GB planes are seen to distinctly facet into symmetric \{$\bar{7}520$\} and asymmetric \{$10\bar{1}0$\} // \{$11\bar{2}0$\} segments of 20-50~nm length. Additionally, EDS reveals preferential segregation of iron (Fe) in every alternate symmetric \{$\bar{7}520$\} segment. Both the faceting and the segregation are explained by a difference in the CSL density between the facet planes. Furthermore, in the GB plane containing Fe segregation, atom probe tomography is used to experimentally determine the GB excess solute to be 1.25~atoms/nm$^{2}$. In summary, the study reveals for the first time a methodology to obtain bicrystalline Ti thin films with strong faceting and an anisotropy in iron (Fe) segregation behaviour within the same family of planes.

\end{abstract}

\maketitle

\section{Introduction}

Titanium (Ti) and its alloys have a high strength-to-weight ratio, excellent corrosion resistance and biocompatibility \cite{BOYER1996}. These properties make it an attractive structural material, mainly in the aerospace and biomedical industries \cite{Banerjee2013}. Despite possessing the majority of the desired properties for automotive industries, the most significant impediment to Ti alloys' widespread use has been their high cost. Many alloying additions have been investigated in recent decades to not only improve mechanical and high-temperature properties but also to make them affordable \cite{Leyens2003, SBanerjee2007}.

Ti exhibits an allotropic transition from the low-temperature hexagonal close-packed (hcp) $\alpha$-phase to the high-temperature body-centered cubic (bcc) $\beta$ phase at 882$^{\circ}$. The alloying elements added to Ti are classified as $\alpha$ stabilizers (Al, Sn, Ga, Zr, C, O, and N), $\beta$-stabilizers (Fe, V, Mo, Nb, Ta, and Cr), or neutral elements (Zr, Sn, and Si), based on the phase they stabilize. Iron (Fe) is one of the most cost-effective $\beta$ stabilizing alloying elements and has therefore attracted plentiful attention \cite{Pan2021, Peng2020, Reverte2020, Devaraj2016}. The addition of Fe can result in its segregation at grain boundaries (GBs), $\beta$-phase formation, or precipitation of intermetallic compounds. Ti-Fe alloys are either $\beta$-phase or a mixture of $\alpha$- and $\beta$-phase, where the phase-fraction depends on the alloy composition \cite{blasius1976}. When added in excess of 2.5 at.\%, Fe is known to form $\beta$-flakes at the GBs which are deleterious to the material properties \cite{Levi1989}. Additionally, two metastable phases, $\alpha^\prime$-Ti (hcp), and $\omega$-Ti(Fe), have been observed during martensitic transformation of Ti-Fe alloys \cite{Straumal2015, Straumal2018, Kilametov2017, Kilmametov2018, Kriegel2019}. The $\omega$-Ti(Fe) is a high-pressure Ti phase which is retained at low pressure in Ti-Fe alloys. However, the role of Fe in the $\alpha \rightarrow \beta$ and in $\beta \rightarrow \omega$ phase transition is not clear yet \cite{Kriegel2020}. 

A change in the chemical composition of the interfacial region changes the thermodynamic driving force for solid-state phase transition \cite{ZHAO2018318}. Fe segregation at GBs in Ti alloys can also lead to the formation of TiFe or Ti$_2$Fe intermetallic compounds \cite{Illarionov_2020}. These binary intermetallics have been considered as a potential choice for solid-state hydrogen storage applications \cite{Sujan2020}. %The $\beta$-phase in Ti alloys improves its corrosion resistance, fracture toughness and yield strength but reduces the ductility \cite{liu2006design}. The role of alloying elements, especially Fe, in the nucleation and growth of $\beta$-Ti is of a great commercial interest. 
Likewise, Ti-Fe alloys have also been extensively used to fabricate near-net shape parts using the blended elemental powder metallurgy (BEPM) route to achieve cost reduction \cite{Savvakin2012, Raynova2021effect}.
%Attempts to partially or completely replace V or other expensive $\beta$-stabilizers with Fe have been successfully undertaken \cite{BODUNRIN2020139622, Esteban2011}. The addition of Fe in the powder blend enhances the sinterability of Ti alloys because of its high diffusivity \cite{liu2006design, CARMAN20111686}. 
Overall, a multitude of phase transitions have been realized in Ti-Fe alloys but numerous questions on how Fe influences these phase transformations remain unanswered. Some of the answers are likely to lie in the GB segregation of Fe that leads to precipitation or other phase transitions.

In the case of $\alpha$-Ti, the maximum solubility of Fe is less than 0.05~at.\% \cite{Matyka1979, raub1967alpha}. Consequently, when added in excess of solubility limit, Fe must either form secondary phases or segregate at the GBs. Although extensive studies have been performed on GB segregation in many fcc and bcc metals and alloys, limited reports have discussed GB segregation in Ti or other hcp metals. Oxygen and Carbon have been shown to weakly segregate at the Ti GBs although oxygen has a high solubility in $\alpha$-Ti \cite{Aksyonov2017}. In $\beta$-Ti, Fe and Cr have been shown to segregate at the GBs and act as grain refiners when cooling \cite{bermingham2009}. The segregation of Fe at Ti GBs has repeatedly been suggested to be responsible for pinning the GBs, leading to remarkably stable nanocrystalline Ti \cite{Aksyonov2017, Simonelli2020}. Recently, density functional theory (DFT) calculations also confirmed the presence of a high driving force for Fe to segregate at $\alpha$-Ti GB \cite{Aksyonov2017}. However, it is unclear how Fe pins the GBs and whether or not secondary phases contribute to this. In many cubic materials segregation is observed to strongly vary based on the nature of the GB \cite{Zhou2016a}. Nevertheless, no anisotropy in segregation with respect to the GB type in Ti or other hcp metals has ever been reported either experimentally or theoretically.

Such a systematic study of the influence of GB character on material properties requires a template-based approach to obtain desired GBs. For many metals, bicrystalline samples with predefined GB parameters have been grown using the vertical Bridgman technique for specific GB property studies \cite{SCHWARZ2001392}. However, in Ti, the hcp-bcc allotropic transition makes it impossible to fabricate bicrystals to form specific desired GBs. Therefore, a novel thin film deposition route to obtain bicrystalline Ti has been established here. Similar bicrystalline thin films have been demonstrated earlier for other metals like Cu and Au \cite{Dahmen1991, Meiners2020, Westmacott2001, Dehm2005}. To deposit such films, a thorough understanding of the impact of various deposition parameters on the microstructure of the film is required. In physical vapour deposition, the degree of ionisation of the plasma particles determines the ion flux towards the growing film \cite{Sarakinos2007}. A high ion fraction in a discharge is achieved by promoting the electron impact ionization which is achieved by using plasma of high electron density and higher temperature. Such a plasma is formed by using a pulsing unit at the target. Textured Ti films with (0002) out-of-plane orientation were recently reported using a similar deposition route \cite{Ma2019}.

In the following sections, firstly a template-based approach to obtain $\Sigma$ 13 GBs in thin films of Ti on SrTiO$_{3}$ substrate is established. Pulsed magnetron sputtering using a commercially pure Ti target with trace Fe impurities is used to obtain a bicrystalline thin film with columnar grains. Subsequently, electron backscatter diffraction (EBSD) in a scanning electron microscope (SEM) is used to characterize the thin film microstructure. The GBs contained in the thin film are analyzed in detail using scanning transmission electron microscope (STEM) and are observed to be faceted into symmetric and asymmetric segments. Using high resolution energy dispersive spectroscopy (EDS) analysis, the segregation of Fe to these topographically complex GBs is explored.

\section{Experiment details}
\subsection{Thin film deposition}
\begin{figure}[ht!]
    \centering
    \includegraphics[width=0.9\linewidth]{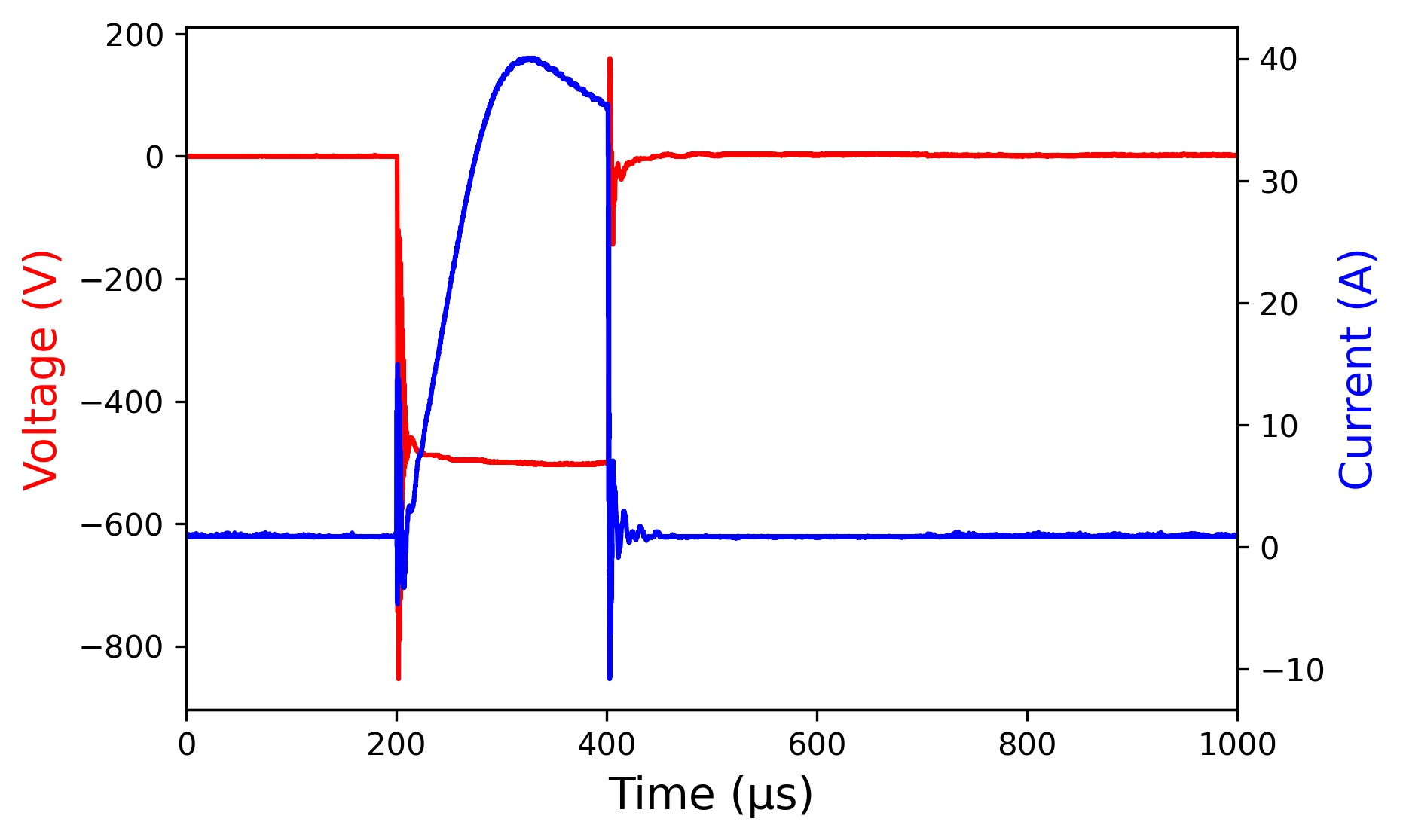}
    \caption{Evolution of the target voltage and current during a single cycle of pulsed magnetron sputtering. A strong voltage overshoot is observed during ignition followed by a steady state of voltage for 200~$\mu$s resulting in a peak target power of 20~kW and peak target power density of  46~W cm$^{-2}$. The pulse period was 2~ms.}
    \label{fig:VI_curve}
\end{figure}

Thin films of Ti were deposited onto 10 $\times$ 10~mm$^2$ SrTiO$_3$ (001) substrates (Crystal GmbH, Germany) using pulsed magnetron sputtering in a commercial deposition system (Ceme Con AG CC 800-9). A rectangular 500 $\times$ 88 $\times$ 10~mm$^{3}$ Ti target of above 99 wt.\% nominal purity (grade 2) with 0.2~wt.\% Fe (0.17~at.\%), 0.18~wt.\% O (0.54~at.\%) and 0.1~wt.\%C (0.54~at.\%) as a major impurity was positioned at a distance of 10~cm to the substrate holder. The base pressure was 2.2 $\times$ 10$^{-6}$~mbar and increased to 3 $\times$ 10$^{-6}$~mbar after heating the substrate to 600$^{\circ}$C using radiation heaters. During deposition, an Ar flow rate was set to 200~sccm leading to a working pressure of approximately 3.8 $\times$ 10$^{-4}$~mbar. A Melec SIPP2000USB-16-500-5 power supply was used and the substrates were at floating potential. Four different deposition conditions are discussed in the following article. The first film was deposited using DC Sputtering with 250~W power for 2.5~h where a deposition rate of 1.33~\AA/s was obtained. The pulsing unit was not used in this deposition. For the subsequent three films, pulsed magnetron sputtering was used. In pulsed magnetron sputtering, the conventional sputtering source is used in a pulsing mode with a predetermined pulse duration ranging from 1~$\mu$s to 1~s to increase the current density. This additional degree of freedom to adjust the pulse duration can be used to tailor desired microstructures \cite{Konstantinidis2006, Schneider2000}. The target voltage and current variation during a single pulse is shown in \cref{fig:VI_curve}. The time average power was set to 1500~W with a pulse duration of t$_{on}$/t$_{off}$ of 200/1800~$\mu$s. This led to the peak current (I$_{ion}$) of 40~A resulting in a dense film and a peak target power density of 46~W cm$^{-2}$. The three films were grown under identical conditions by keeping all the film deposition parameters unchanged but changing the post-deposition annealing duration between 2h, 4h and 8h at 600$^{\circ}$C to investigate its influence on the film microstructure, texture and grain size. The annealing was performed immediately post turning off the plasma without breaking the vacuum because Ti is known to be highly prone to oxidation.  With a deposition rate of about 8.3~\AA/s, a film thickness of 1.5~$\mu$m was obtained in $\sim$30~min. The venting temperature was always \textless 70$^{\circ}$ to reduce surface oxidation \cite{Greczynski2016}.

\subsection{Microstructural characterization}
The preliminary investigation was carried out using a light optical microscope (LOM) to check for any cracks/ defects on the surface. Subsequently, orientation of all grains was mapped using EBSD in an SEM. An in-plane lift-out technique in a dual-beam focused ion beam (FIB) instrument (Thermo Fisher Scientific Scios 2 HiVac) with Ga$^{+}$-ion source was used to extract a transmission electron microscopy (TEM) lamella. The beam current was gradually reduced in several steps starting from 1~nA at 30~kV for coarse milling to eventually 27~pA at 2~kV for final polishing to obtain a thickness of \textless 100~nm. Probe corrected STEM in a Titan Themis 80-300 (Thermo Fischer Scientific), was used at an acceleration voltage of 300kV. A semi-convergence angle of 23.8~mrad was used for imaging. With a camera length of 100~mm, collection angles of 78-200~mrad and 38-77~mrad were obtained for the high-angle annular dark field (HAADF) and the annular dark field (ADF) detectors, respectively. Thermo Scientific ChemiSTEM Technology using four in-column Super-X detectors was used with a beam current of $\sim$50~pA for the EDS analysis. Laser-pulsed atom probe tomography (APT) was performed in a LEAP$^{TM}$ 5108XR (CAMECA) at a repetition rate of 200 kHz, a specimen temperature of about 50 K, a pressure lower than $1 \times 10^{−10}$~Torr ($1.33 \times 10^{−8}$~Pa) and a laser pulse energy of 20~pJ. The evaporation rate of the specimen was 5 atoms per 1000 pulses. Datasets were reconstructed and analyzed with the AP suite 6.1 software based on the voltage curves.

Using the results from APT, the interfacial excess was experimentally determined by selecting a region of interest (ROI) across the interface where the solute is segregated and plotting the so-called, \textit{ladder diagram}. A \textit{ladder diagram} is established by taking the total number of atoms in the ordinate and the integral of solute atoms in the abscissa. A linear fit within the concentration profile of the two grains can be extrapolated to the Gibbs dividing surface (GB surface) to find the solute content in both the grains, N$_a$ and N$_b$. Using this, the Gibbsian interfacial excess is calculated as:
\begin{equation}
    \Gamma_{excess} = \frac{N_{b} - N_{a}}{(Detection\: efficiency\: \times \:Area\: of\: ROI)}
    \label{eq:GBExcess}
\end{equation}
Additional details are described in \cite{Krakauer1993}. The detection efficiency was 0.52.

\section{Results and Interpretation}

\subsection{Evolution of thin film microstructure and grain boundaries}

\begin{figure*}
    \centering
    \includegraphics[width=0.7\linewidth]{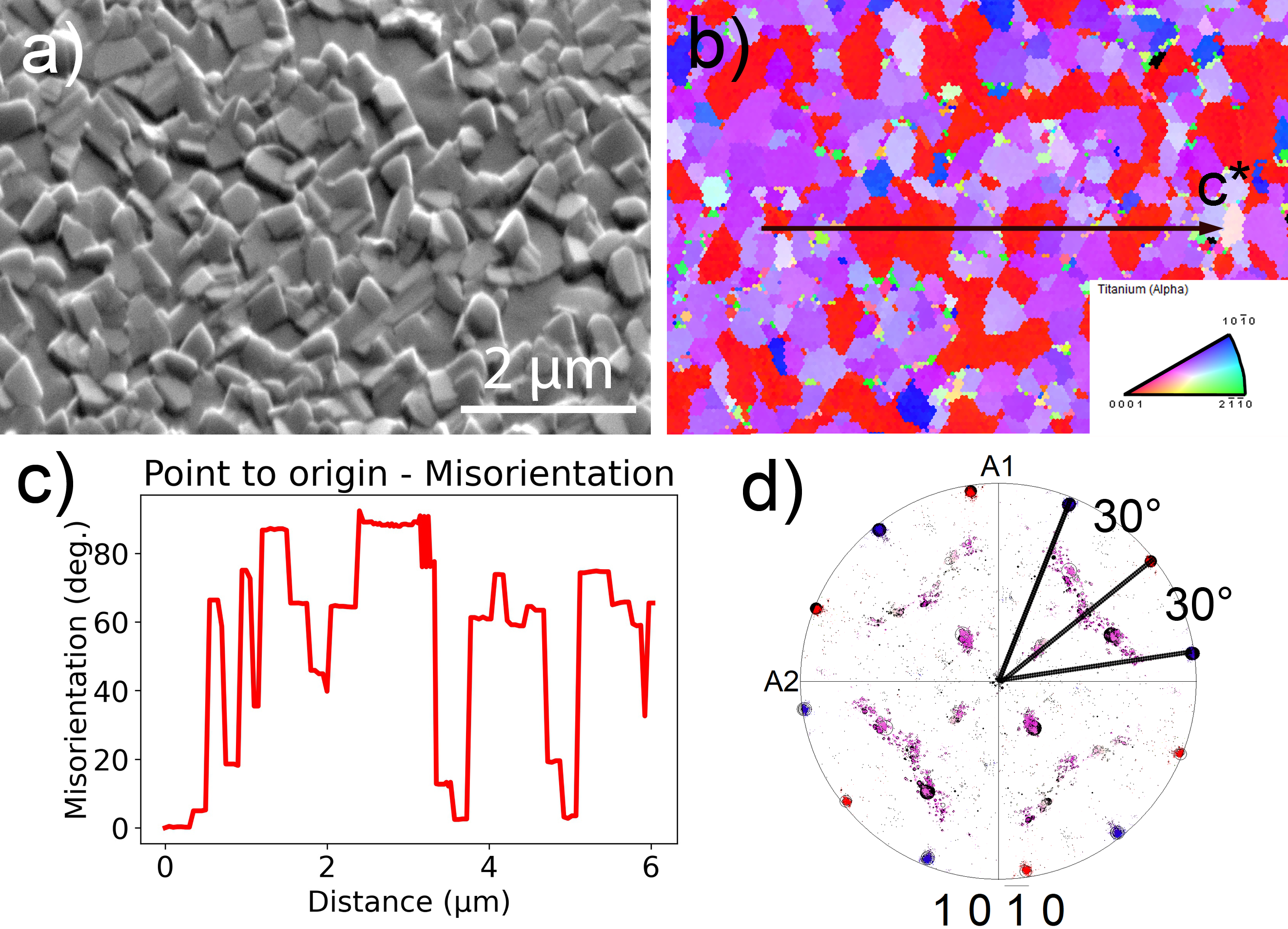}
    \caption{Ti film deposited using DC Sputtering with 250~W power on SrTiO$_3$ at 600$^{\circ}$C and post annealed at 600$^{\circ}$C for 2~h. a) SE image showing rough surface and small grain size. b) Inverse pole figure (IPF) map obtained from electron backscatter diffraction (EBSD) revealing the two dominant surface plane orientations to be $\{10\bar{1}1\}$ and $(0002)$. The arrow highlights the points along which a point-to-origin misorientation profile chart is plotted in (c). d) Pole figure confirming the dominance of two surface plane orientations with a fiber texture of $\{10\bar{1}1\}$ planes and only two in-plane rotations of $(0002)$ planes.}
    \label{fig:film1}
\end{figure*}

\begin{figure*}[htp]
    \centering
    \includegraphics[width=0.7\linewidth, keepaspectratio]{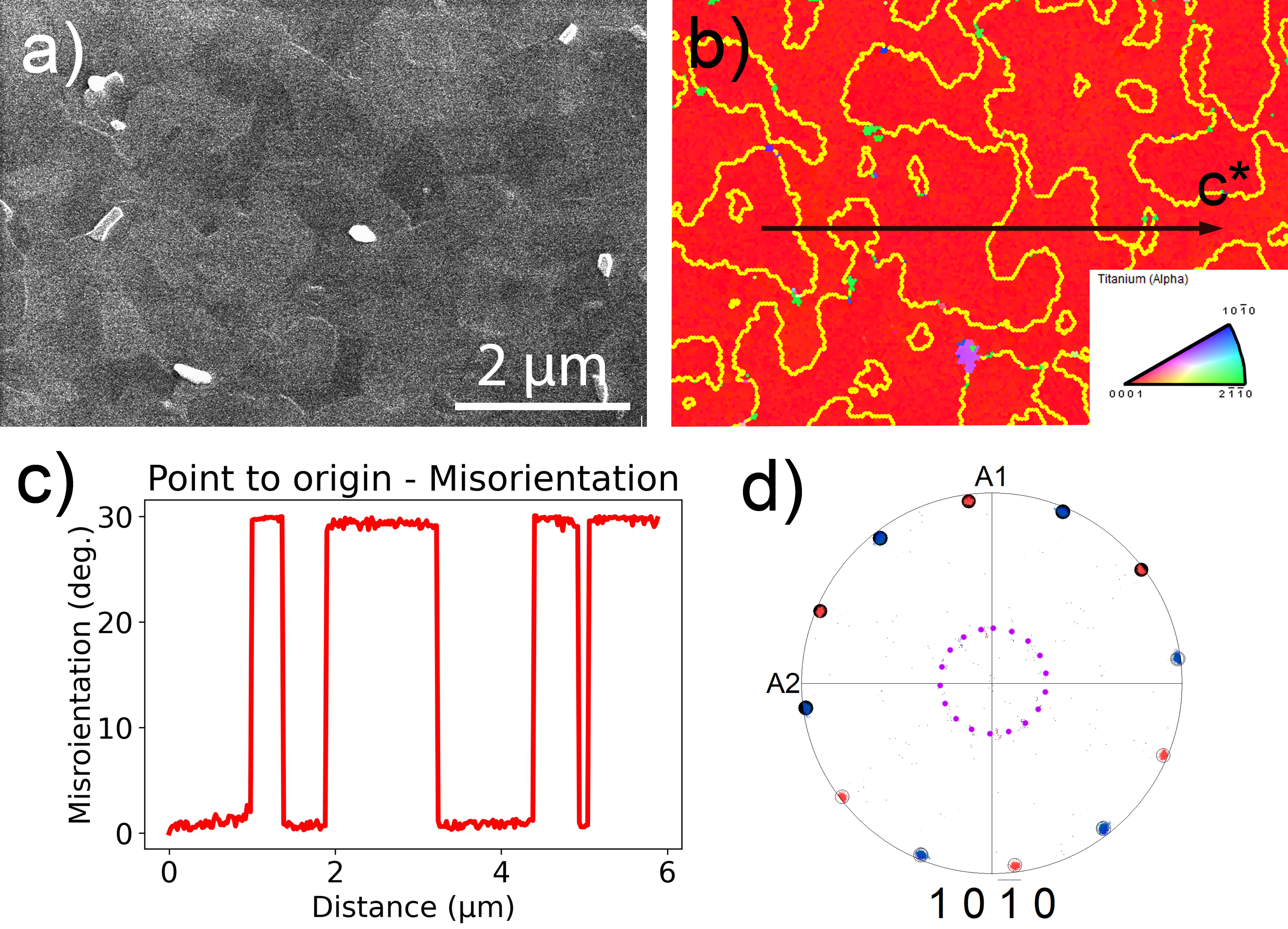}
    \caption{a) SE image of Ti film deposited on SrTiO$_3$ at 600$^{\circ}$C using pulsed magnetron sputtering with a pulse duration of 200~$\mu$s and 1500~W power followed by post annealed at 600$^{\circ}$C for 4~h showing smoother surface and relatively larger grain size. b) Inverse pole figure (IPF) map obtained from EBSD with $\Sigma13$ grain boundaries highlighted in yellow. The arrow highlights the points along which a point-to-origin misorientation profile chart is plotted in (c). d) Pole figure obtained from IPF confirming the $\sim30$ $^{\circ}$ misorientation corresponding to $\Sigma13$ [0001] grain boundaries with the orientations highlighted in blue and red, respectively. A dashed purple circle highlights the presence of trace $10\bar{1}0$ orientated grains.}
    \label{fig:film2}
\end{figure*}

\begin{figure*}[htp]
    \centering
    \includegraphics[width=0.8\linewidth,height=\textheight, keepaspectratio]{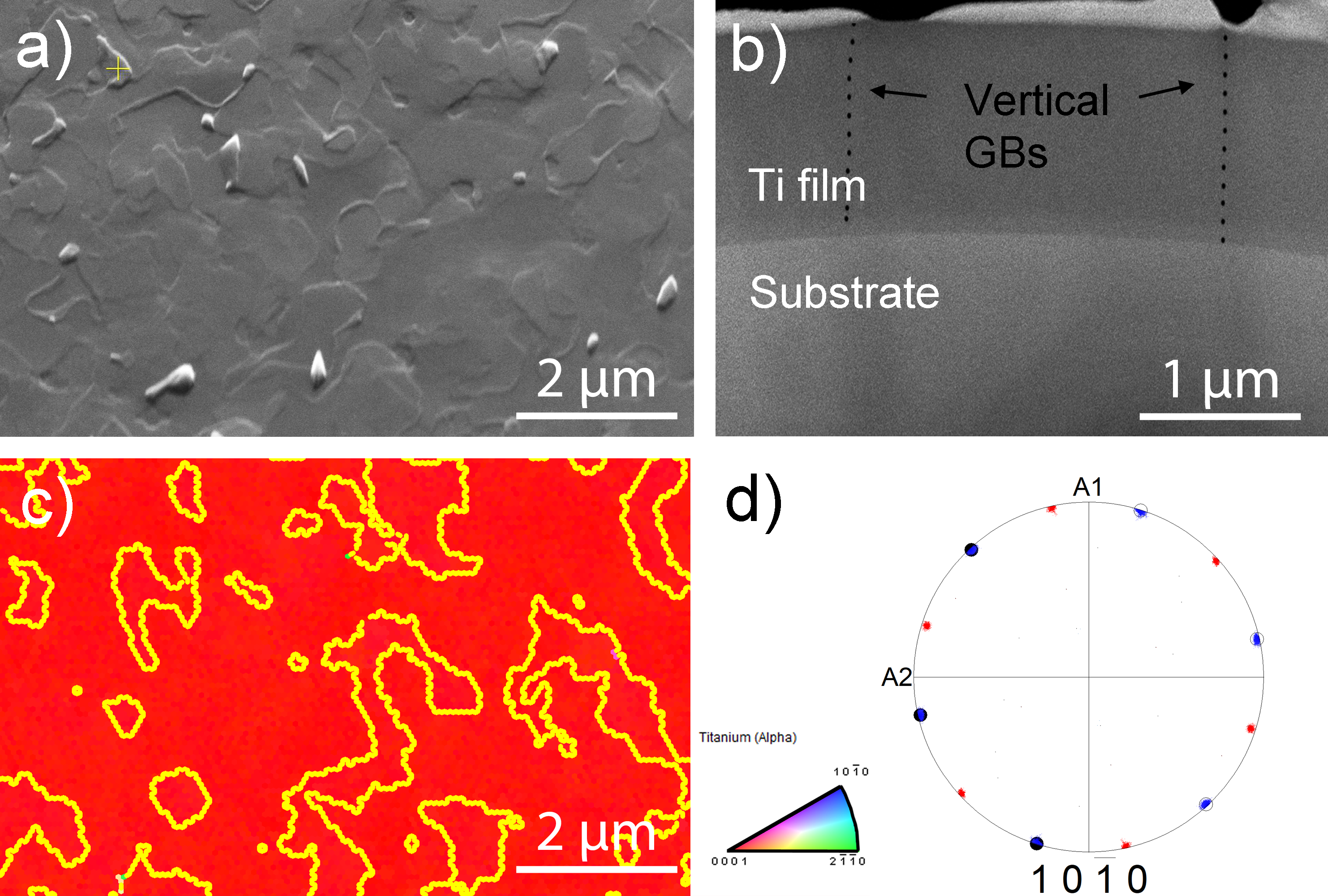}
    \caption{Ti deposited on SrTiO$_{3}$ at 600$^{\circ}$C and post annealed at 600$^{\circ}$C for 8~h a) SE image showing traces of grain boundaries. b) Cross-section image taken in SEM-STEM confirming columnar grains. c) Inverse pole figure (IPF) map obtained from EBSD with $\Sigma13$ grain boundaries highlighted. d) Pole figure obtained from IPF confirming the $\sim$30$^{\circ}$ misorientation corresponding to $\Sigma13$ [0001] grain boundaries.}
    \label{fig:film3}
\end{figure*}

\begin{figure*}[t]
    \centering
    \includegraphics[width=0.8\linewidth,height=\textheight, keepaspectratio]{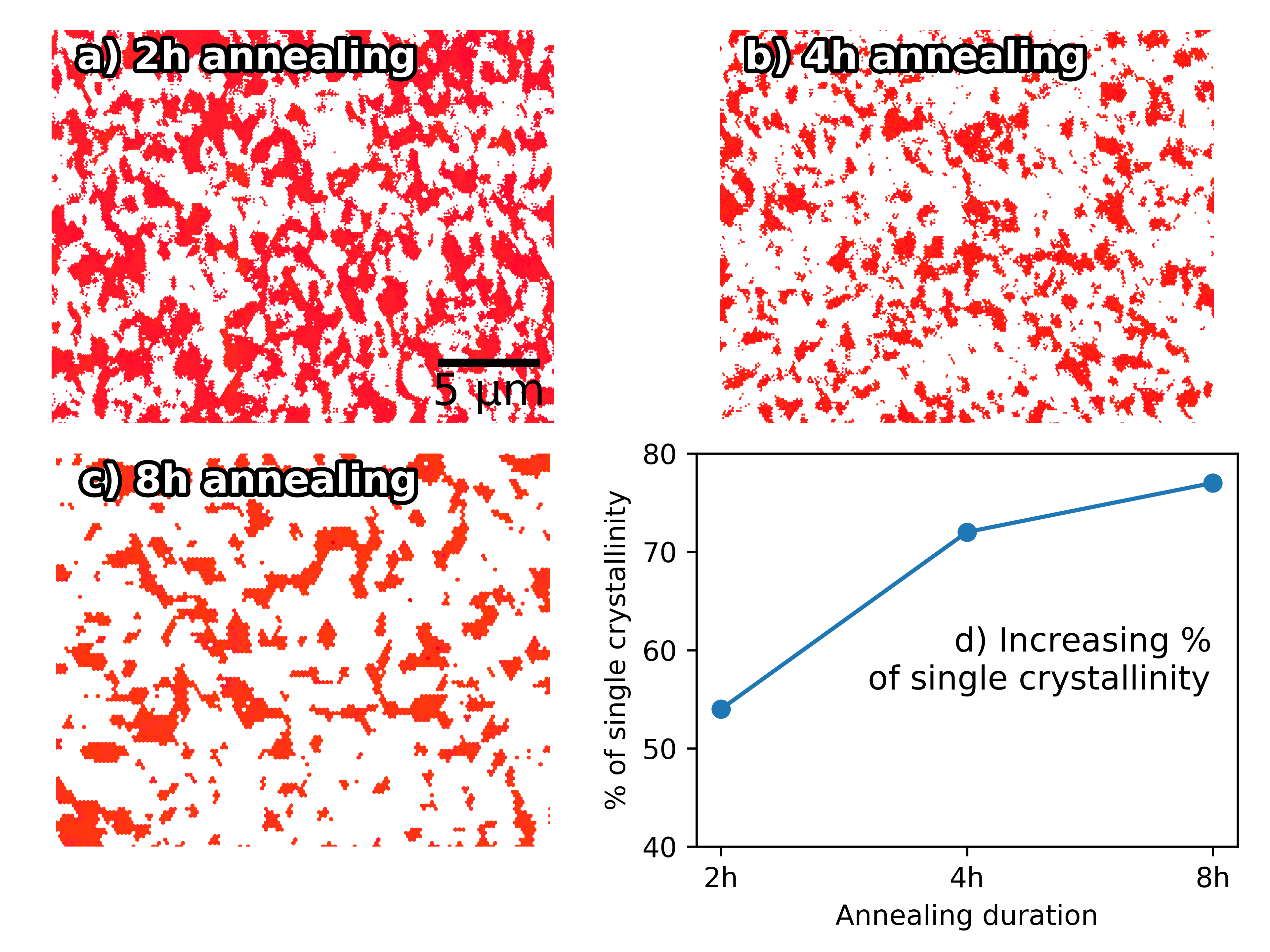}
    \caption{Partition of the orientation map of Ti deposited on SrTiO$_{3}$ at 600$^{\circ}$C highlighting the decreasing fraction of one orientation at the expense of other. a), b) and c) are the films annealed at 600$^{\circ}$C for 2h, 4h and 8h respectively. d) The increasing area fraction of G1 orientation plotted with respect to annealing duration demonstrating grain growth and increasing \% of single crystallinity.}
    \label{fig:ggrowth}
\end{figure*}

The Ti film deposited on SrTiO$_3$ (001) at 600$^{\circ}$C using DC sputtering is characterized by SEM and EBSD as shown in \cref{fig:film1}. Following the deposition, the film was post-annealed at 600$^{\circ}$C for 2~h in the deposition chamber. The secondary electron (SE) image in \cref{fig:film1} a) reveals a rough surface and small grains. \cref{fig:film1} b) shows the crystallographic orientation map based on the [0001] inverse pole figure (IPF) obtained using EBSD. The grain size is measured to be $\sim$500~nm using the line intercept method \cite{hilliard1964}. To visualize the change in misorientations, a black line is highlighted inside the orientation map in \cref{fig:film1} b). The misorientation between every point on the line and the first point (origin) on the line is displayed as a misorientation profile chart. In the profile, a range of varying orientations is observed in \cref{fig:film1} c). Using the $[10\bar{1}0]$ pole figure in \cref{fig:film1} d), two dominant textures are observed. First, a strong $(10\bar{1}1)$ fiber texture is seen revealing the presence of all possible in-plane rotations. These grains are highlighted in purple in both the orientation map and the pole figure. Second, a $(0002)$ texture is observed with only two in-plane grain rotations, with each orientation highlighted in red and blue in the pole figure. The pole figure comprises a single point per grain that is weighted by grain size; thus, the distribution of both orientations can be seen to be approximately equal. It is known that the (0002) plane has the lowest surface energy in Ti due to the highest atomic density and $(10\bar{1}1)$ has the least strain energy due to the lowest elastic modulus \cite{Checchetto1997}. Hence, the two orientations seem to compete and both are observed by EBSD in the film deposited by DC Sputtering.

To understand the influence of higher ionization of the plasma and increased adatom mobility on the film microstructure, pulsed magnetron sputtering was subsequently used to deposit three additional films. A Ti film was deposited using pulsed magnetron sputtering at 600$^{\circ}$C and post-annealed at the same temperature for 2~h. A larger grain size (see \cref{Tab:Tab1}) is seen for the film deposited by pulsed magnetron sputtering compared to the film deposited using DC sputtering. Likewise, for the film annealed for 4~h, a much smoother surface and larger grain size is observed, as seen in \cref{fig:film2}. A cross-section FIB lamella was prepared to resolve the microstructure of the film in growth direction. The majority of the grains were found to be columnar. In \cref{fig:film2} b), the orientation map obtained from EBSD confirms that almost all of the grains have a [0001] surface plane normal. The pole figure shown in the \cref{fig:film2} d) is obtained from the same data set. It confirms the $\sim30$ $^{\circ}$ misorientation corresponding to $\Sigma13$ [0001] GBs. Furthermore, as previously observed, all of the grains with (0002) surface plane normal have either of the two orientations shown in the pole figure, which are denoted by red and blue. As a result, the film has a mazed bicrystalline microstructure, as seen in the IPF. The same is also confirmed by the presence of only two alternating orientations in the misorientation profile chart in \cref{fig:film2} c). A purple dashed-circle is drawn in \cref{fig:film2} d) to point at the additional reflections at $\sim$25$^{\circ}$ away from the center. As every reflection in the pole figure is weighted by the grain size, these additional reflections correspond to small grains of other orientations. The high peak current of 40~A during pulsed magnetron sputtering in contrast to 5~A during DC sputtering is responsible for the shift of surface plane orientation from $(10\bar{1}1)$ and $(0002)$ in the DC sputtered film to all the grains being only $(0002)$ oriented in the pulsed magnetron sputtered film. 

\begin{figure*}[ht!]
    \centering
    \includegraphics[width=0.7\linewidth,height=\textheight, keepaspectratio]{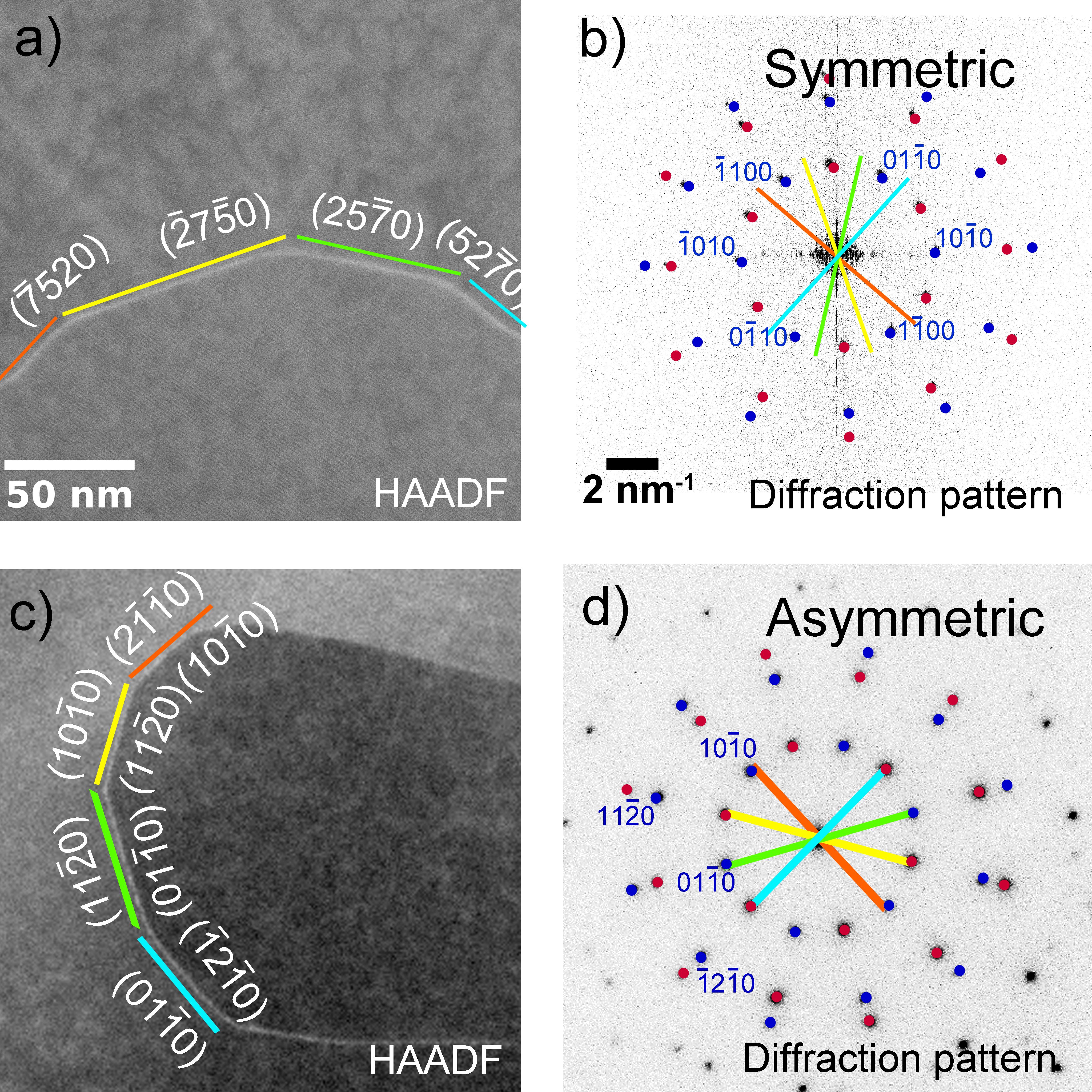}
    \caption{a) HAADF image of $\Sigma$13 [0001] Ti GB in the pulse magnetron sputtered film that was post-annealed for 8~h showing GBs faceted such that all GB planes are symmetric. b) Colour inverted diffraction pattern acquired using TEM from the same region as shown in (a) to find GB planes. Diffraction spots from both the grains are indicated using red and blue circles, and each GB plane is represented by a different colour. The indexing of the planes confirms the symmetric faceting. c) HAADF image of same film in another region showing asymmetric GB facets. d) Diffraction pattern obtained from the same region as (c) confirms the asymmetric facets.}
    \label{fig:Facet}
\end{figure*}

\begin{figure*}[ht]
    \centering
    \includegraphics[width=0.7\linewidth,height=\textheight, keepaspectratio]{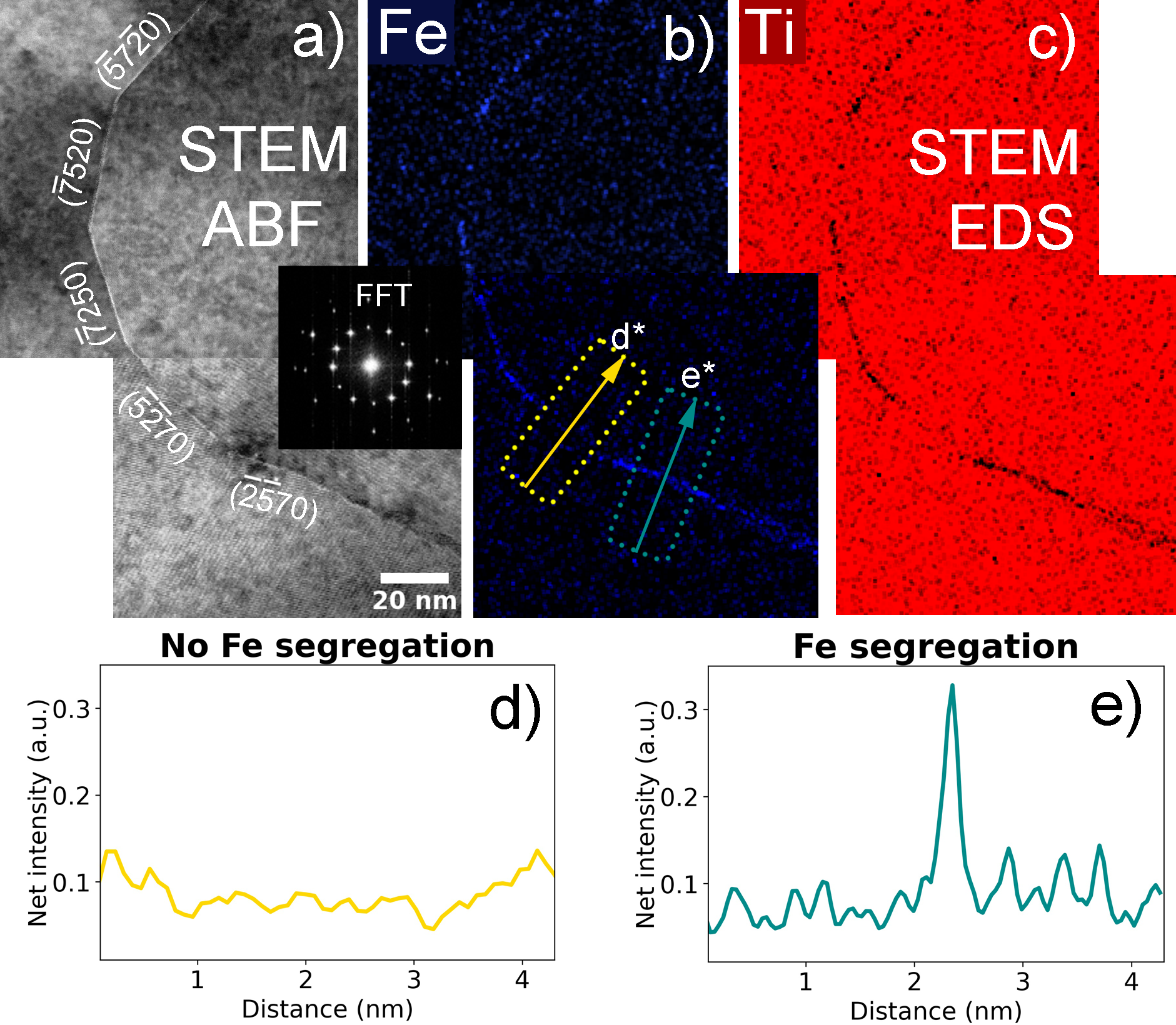}
    \caption{a) STEM annular-BF image showing GB facets with symmetric \{${\Bar{7}520}$\} GB planes. EDS reveals segregation of b) Fe in alternate GB facet, c) depletion of Ti in the Fe rich region. The Fe rich GBs are found to be planes with lower CSL density. d), e) Net-intensity line profile across the two symmetric facets revealing a strong segregation of Fe in only the $(\bar{2}\bar{5}70)$ facet.}
    \label{fig:Fe_seg_eds}
\end{figure*}

When the post annealing time is extended to 8~h, the film surface remains smooth and a further increase in grain size to 3.76 $\pm$ 2.4~$\mu$m is measured using EBSD, as seen in \cref{fig:film3}. The increase in grain size with respect to the annealing duration is tabulated in \cref{Tab:Tab1}. EBSD also verifies that the film continues to show the same mazed bicrystalline microstructure as observed previously. When comparing the films annealed for 4~h and 8~h, the isolated grains of other orientations are seemingly overgrown by the larger grains having $(0002)$ orientation in the 8~h sample. The novelty of both the pulsed magnetron sputtered films is that, not only they are bicrystalline and thus strongly textured but also all the grains are completely columnar from the substrate to the film surface, as seen in the SEM-STEM image of a cross-section of the film in \cref{fig:film3} b). Such textured films are of great interest, for example in microelectronics for applications such as diffusion barriers in integrated circuits \cite{Chen2002}. In the present context, the columnar grains are necessary to ensure that the GBs are edge-on in plan-view. The pole figure in \cref{fig:film3} d) again shows the presence of only $(0002)$ out-of-plane orientations as discussed previously. After 8~h of annealing, no additional fringe orientations are observed.

The orientation maps show that one of the two (0002) orientations present is a continuous large grain that extends all over the substrate. This orientation is here on wards referred to as 'G1'. The other orientation, 'G2', is present as small islands surrounded by the G1 grain. This is the typical microstructure observed in other bicrystalline films such as (110) Au film on (001) Ge substrate \cite{Radetic2012}. Using the TSL-OIM software, a partition of only G2 orientation is created from the acquired EBSD data for all three films deposited using pulsed magnetron sputtering. The G2 oriented grains are highlighted in the \cref{fig:ggrowth} a), b) and c) for 2~h, 4~h and 8~h of annealing, respectively. As a function of annealing time, the G1 orientation grows at the expense of G2 orientation resulting in an increase in the single crystallinity percentage, as shown in \cref{fig:ggrowth} d).

\begin{table}
\caption{Deposition conditions and grain size of all films}
\begin{tabular}{cll}
Deposition method                & Annealing T \& t         &   Grain size ($\mu$m)\\
\hline
DC sputtering                    & 600$^{\circ}$C \& 2~h    &   0.48 $\pm$ 0.2 \\
Pulsed magnetron sputtering      & 600$^{\circ}$C \& 2~h    &   2.06 $\pm$ 0.9 \\
Pulsed magnetron sputtering      & 600$^{\circ}$C \& 4~h    &   2.43 $\pm$ 1.4 \\
Pulsed magnetron sputtering      & 600$^{\circ}$C \& 8~h    &   3.76 $\pm$ 2.4 \\
\end{tabular}
\label{Tab:Tab1}
\end{table}

\subsection{Grain boundary faceting}

The $\Sigma13$ GBs in \cref{fig:film2} b) and \cref{fig:film3} c) are observed to be continuously curved and to form a mazed bicrystalline microstructure. Such GB curvature is often accommodated by faceting if the inclinational dependence of the grain boundary energies is anisotropic \cite{Dahmen1991, Straumal2015}. Upon investigation at a higher magnification, the GBs are observed to be faceted, as seen in \cref{fig:Facet} a). The plane normal of each facet plane is 30$^\circ$ apart from each other. A selected area diffraction pattern is obtained from the same region using TEM to index the GB planes. Because the GBs are all edge-on, as previously demonstrated in \cref{fig:film3} b), the GB planes can be indexed by simply locating the intersection of the GB plane normals to the great-circle in the (0001) sterographic projection of Ti. Using this, the GB facets are indexed to be $\{\bar{7}520\}$ symmetric planes. From the diffraction pattern it is also apparent that \{${\Bar{7}520}$\} is the only conceivable symmetric GB plane family in $\Sigma13$ [0001] hcp GBs. GB planes with any other Miller indices would be asymmetric. A preference for the symmetric GB planes is readily apparent. 
In another instance, as seen in \cref{fig:Facet} c), although the GB facet normals are still 30$^{\circ}$ apart, the planes were indexed to be low-index asymmetric $\{10\Bar{1}0\}$ // $\{2\bar{1}\bar{1}0\}$ planes. The diffraction pattern shown in the \cref{fig:Facet} d) confirms the asymmetricity of the facets.
According to authors' observations, out of over $\sim$200~$\mu$m of GB less than $\sim$30~$\mu$m of the GB is asymmetric, which corresponds to less than $\sim$15\% of the GB being asymmetric.

\subsection{Anisotropic Fe segregation in symmetric GB facets}

Fe was present as an impurity in the sputtering target therefore it is of interest to examine where it is present in the deposited films. In the film deposited using pulse magnetron sputtering and post-annealing for 8h at 600$^{\circ}$, the $\Sigma$13 [0001] Ti GB is seen to be composed of \{${\Bar{7}520}$\} symmetric GB facets, as shown in \cref{fig:Fe_seg_eds} a). The facets are similar to the GB faceting discussed in \cref{fig:Facet}. The elemental distribution map of Fe, acquired using STEM-EDS at 300kV, reveals a preferential segregation of Fe to every alternate symmetric facet. In \cref{fig:Fe_seg_eds}, Fe segregation is observed at $(\bar{5}7\bar{2}0)$, $(\bar{7}250)$, and $(\bar{2}\bar{5}70)$ facets, but Fe is not detected at $(\bar{7}520)$ and $(\bar{5}\bar{2}70)$ facets. The counts of the Fe signal in the EDS spectrum are integrated along the highlighted arrow over the marked region as seen in \cref{fig:Fe_seg_eds} b). \cref{fig:Fe_seg_eds} d) and \cref{fig:Fe_seg_eds} e) show line profiles across the corresponding facets to clearly illustrate the segregation of Fe at the $(\bar{2}570)$ GB plane and the absence of Fe-segregation in the $(\bar{5}\bar{2}70)$ facet. The FWHM of the spacial distribution of Fe in the \cref{fig:Fe_seg_eds} e) is $\sim$0.15~nm confirming the segregation of Fe is limited to the GB.

To quantify the amount of segregation APT was used. Firstly, the GB facet containing Fe was targeted to be lifted-out using the conventional FIB sample preparation technique and field evaporated in laser-mode. After the reconstruction of data using AP Suite, the atom distribution maps of Ti and O are obtained, as shown in \cref{fig:APT} a). Both of them are seen to be distributed uniformly in both the grains and the GB region. Fe is seen in \cref{fig:APT} b), c) and d) to distinctly segregate to the GB. To delineate the Fe GB segregation, a 0.5~at\% Fe isoconcentration surface is plotted in \cref{fig:APT} b) and as seen in \cref{fig:APT} d), the GB is edge-on. A cylinder of 30~nm diameter and 50~nm length is highlighted as the selected region of interest (ROI). Although the GB extends over the entire cross-section of the APT tip, Fe is seen enriched at only a fraction of the GB area. The composition along the ROI in both \cref{fig:APT} b) and \cref{fig:APT} c) are plotted in \cref{fig:APT} e) and \cref{fig:APT} f), respectively. Negligible Fe segregation is observed in \cref{fig:APT} e) whereas significant Fe segregation of up to $\sim$0.5~at.\% is observed in \cref{fig:APT} f). The Fe segregation is distributed to a width of $\sim$8~nm (FWHM \textless 4~nm). The segregation width is significantly larger than the expected GB width due to the artefacts from field evaporation and aberrations in the trajectory of ions \cite{Miller1992}. Nevertheless, this apparent increase of GB width is inconsequential for a homo-phase boundary as the concentration on both sides of the Gibbsian dividing surface can be considered to be identical. Although not highlighted in the \cref{fig:APT} b), c) and d) for clarity, three ROIs of the same dimensions were taken for better statistics. As seen in the \cref{fig:APT} g), plotting the number of Fe atoms against the total number of atoms of all elements in the region of interest allows us to measure the  number of GB excess Fe atoms per unit area of the interface, N$_{Fe}$. Using the equation \cref{eq:GBExcess}, $\Gamma_{Fe}$ is found to be 1.25 $\pm$ 0.1~atoms/nm$^2$. Since the \{${\Bar{7}250}$\} GB has $\sim$8-10~at/nm$^{2}$ (ambiguity arises because of the dependence of planar-atomic-density on the width of a high-index GB plane), the amount of segregation can also be described as $\sim$0.2 monolayers, assuming that the Fe segregation is limited to the GB plane. The relationship between misorientation and GB excess property measured using APT has been discussed in cubic metals \cite{Zhou2016a, stoffers2015}, however no report for hcp metals was found in the literature. To emphasize that the segregation of Fe is limited to the GB plane, and to show its distribution, an areal density plot of Fe in the GB plane is shown in \cref{fig:APT} h).

\section{Discussion}

\subsection{Thin film deposition}

In a plasma with a large degree of ionization, the ion flux to the substrate is larger than for discharges with a low degree of ionization \cite{Schneider2000}. The magnitude of ion current affects the morphology evolution which is evident when comparing the SE image of the film deposited by DC sputtering in \cref{fig:film1} a) and the SE image of the film deposited by pulsed magnetron sputtering in the \cref{fig:film2} a).
The incoming sputtered atoms from the target when adsorbed on the substrate surface are called adatoms. Due to the dense plasma in pulsed magnetron sputtering, adatom mobility to low surface energy sites with high coordination is promoted. Enhanced surface diffusion eliminates the voids and surface mounds and consequently reduces the surface roughness of the grown films. It also leads to a smoother surface finish \cite{SITTINGER2008}. Increase in surface smoothness and density of films on using higher duty cycles have been reported earlier in TiO$_x$ \cite{Sarakinos2007}. 

Additionally, the lowest surface energy plane, $(0002)$, dominated over the lowest strain energy plane, $\{10\bar{1}1\}$. The high flux of ions due to pulsing leads to the increased momentum transfer between the plasma and the condensed metal atoms. Such a high flux increases the mobility of surface adatoms and accommodates them on planes of the lowest surface energy. A similar change in texture evolution in changing the deposition method from DC sputtering to pulsed magnetron sputtering favouring the high atomic density surface planes has also been reported in other materials \cite{Petrov1993, Hultman1995, SARAKINOS20101661}.

Moreover, all the pulsed magnetron sputtered films were observed to have columnar grains. This can be explained using the structure zone model (SZM). The SZM is commonly used to determine the dependence of film microstructure on the discharge pressure and the homologous temperature \cite{Thornton1974}. The film microstructure according to this model is categorized into four zones, namely Zone 1, Zone T, Zone 2, and Zone 3. Detailed discussion on the model can be found in \cite{kusano2019}. In the present film, the deposition temperature of 600$^{\circ}$C lies in the middle of zone T and zone II (T$_{s}$ / T$_{m}$ $\sim$ 0.4; T$_{s}$: deposition T, T$_{m}$: melting point of target ) which is expected to lead to columnar grains. However, the most important characteristic of these films is that all of the GBs are identified as $\Sigma13$. It must be noted that even with a detailed EBSD scan with a step size of \textless 10~nm, the GBs seem to be rounded, forming a meandering bicrystalline microstructure. Although pulsed magnetron sputtering of Ti has been used earlier to obtain Ti films having similar smooth surface and columnar grains, the GBs were not characterized in these works \cite{Jing2012, Sarakinos2007}.
 
\subsection{Grain boundary faceting}

\begin{figure*}[p!]
    \centering
    \includegraphics[width=0.9\linewidth]{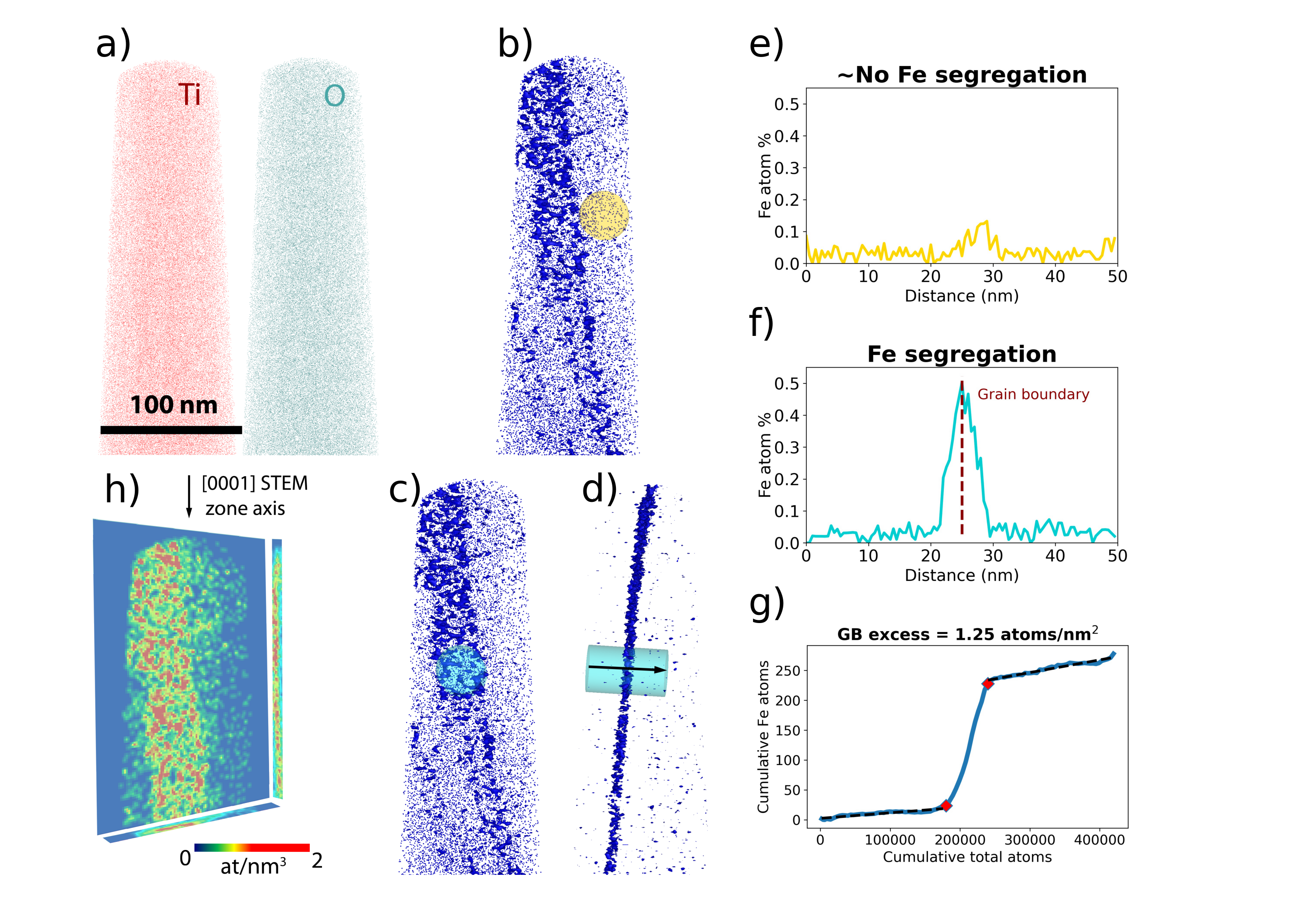}
    \caption{a) Reconstruction point cloud from atom probe tomography (APT) showing Ti and O atoms marked in red and green, respectively. No enrichment or depletion of either of the elements is observed at the GB. b) Isocomposition surface of 0.5~at.\% Fe delineates the GB and demonstrates the segregation limited to a fraction of the GB area present in the tip. The region of interest (ROI) used for composition analysis through the GB segment having no Fe segregation is highlighted (in gold). c), d) ROI passing through the Fe-enriched GB region is depicted in both front and side view, respectively (in teal). e), f) The composition profile of Fe plotted against the length of the highlighted ROI exhibits $\sim$no Fe segregation and Fe-segregation, respectively. The Fe-segregation is limited to GB width and also limited to only a fraction of the GB. g) The ladder diagram shows GB segregation of Fe with equally low solubility in the grain interior on both sides of the interface; the Gibbsian GB excess is calculated as explained in \cite{Krakauer1993}. h) The in-plane atomic density distribution of Fe at the GB plane.}
    \label{fig:APT}
\end{figure*}

\begin{figure*}[hp!]
    \centering
    \includegraphics[width=0.8\linewidth,height=\textheight, keepaspectratio]{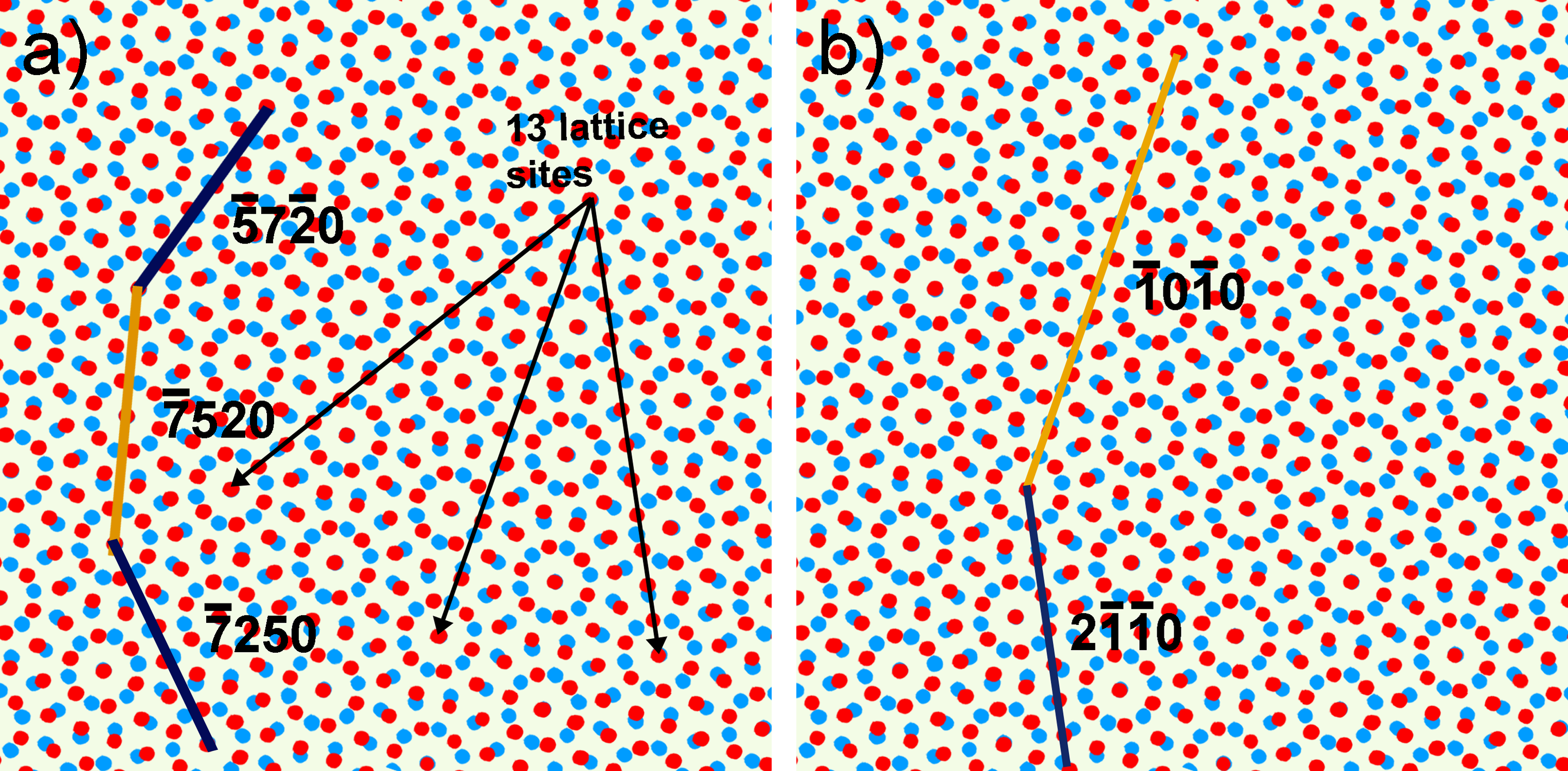}
    \caption{Schematic of $\Sigma13$ GB with planes a) $\{\Bar{7}520\}$ and $\{\Bar{5}7\Bar{2}0\}$ consisting of highest and second highest planar coincident site density (PCSD), respectively leading to reduction of GB energy. b) $\{2\Bar{1}\Bar{1}0\}$ and $\{10\Bar{1}0\}$ have a lower PCSD but higher effective d-spacing (d$_{eff}$) leading to the second most frequently observed GB facet plane.}
    \label{fig:csl}
\end{figure*}

Faceting is the dissociation of a GB into segments with different GB plane inclination but same overall misorientation. It principally occurs to reduce the overall GB energy. The total energy of a faceted GB is the sum of the energy of individual GB segments and their interaction energy at the facet junctions. As faceting leads to an increase in GB plane area, the facets must have lower energy than the parent GB for faceting to occur. Although GB faceting has been widely reported in many cubic metals \cite{BISHOP1971, Balluffi1977, Muschik1993}, to the best of the authors' knowledge, no experimental evidence of GB faceting has been reported for Ti. Here, we show that the GB plane which would otherwise be asymmetric due to its continuous curvature, dissociates into distinct facets.

Since all the grains are columnar, it is evident that all the GBs in a basal-plane textured film are prismatic in nature. Moreover, there is a competition between the low index asymmetric prismatic $\{10\Bar{1}0\}$ // $\{2\bar{1}\bar{1}0\}$ GB planes and the symmetric prismatic \{${\Bar{7}520}$\} GB planes. The prismatic planes have been shown to be the preferred GB facet plane in other hcp materials \cite{Glowinski2014, KELLY201622}. A 3D-EBSD study on bulk-Ti reported a large fraction of grains with misorientation $\leq$ 30$^\circ$ having prismatic GB planes with a preference for \{$\bar{7}520$\} \cite{KELLY201622}. Therefore, not only are the \{$\bar{7}520$\} GBs preponderant in thin films, but they are also frequently observed in the bulk commercially pure Ti. Also, prismatic planes are preferred in the $\beta$ $\xrightarrow[]{}$ $\alpha$ martensitic phase transformation in Ti \cite{FARABI2018147}. The \{$\bar{7}520$\} prismatic planes are also predominant in $\Sigma13$ [0001] GBs in Mg \cite{OSTAPOVETS2014102}. Using atomistic calculations, Ostapovets et al. were able to show that the minima of [0001] tilt GB energy corresponds to \{$\bar{7}520$\} \cite{OSTAPOVETS2014102}. Although, similar GB energy calculations via MD simulation for Ti [0001] tilt-GBs are missing, it is expected that Ti could follow a similar trend as in Mg. Therefore, the observed GB faceting in \cref{fig:Facet} a) and \cref{fig:Fe_seg_eds} a) into \{$\bar{7}520$\} symmetric segments can be considered to be a result of the GB energy-minimization in hcp materials.

To develop a thermodynamic understanding of the anisotropic segregation of Fe observed in the \cref{fig:Fe_seg_eds}, it is necessary to know the atomic structure and the local energetics of the particular GB. The atomic structure investigation is beyond the scope of the present study but a simpler approach to rationalize the observations can be used. The CSL model is the most widely applied tool for classifying GBs. It distinguishes grain misorientations that place a large fraction of the lattice sites of the two grains in coincidence from the remaining 'general' misorientations. The sigma value represents the inverse of the number of the coinciding sites. If the GB plane passes through the coinciding points, then such a GB plane can have low energy. The term, 'planar coincident site density' (PCSD) is used to quantify the density of CSL sites on the GB plane. In a SrTiO$_{3}$ $\Sigma$3 (111) twist boundary, the high PCSD was held responsible for the observed GB plane \cite{Rohrer2004}. To illustrate the role of PCSD in GB plane selection, a dichromatic pattern of hcp (0001) is drawn in \cref{fig:csl}. The $(\bar{7}520)$ symmetric plane that is highlighted in 'gold' has the highest PCSD, $\Gamma$ = 0.43, in $\Sigma13$ [0001] hcp GBs. This is followed by the $(\bar{5}7\bar{2}0)$ plane highlighted in 'dark-blue' which has a $\Gamma$ = 0.33. Caution must be exercised because both of these planes belong to the same family of planes and should thus be interchangeable. Although the \{${\Bar{7}520}$\} consists of twelve planes, six of them have a lower PCSD than the remaining six in the $\Sigma13$ misorientation. This peculiar behaviour is an outcome of the six-fold symmetry of the basal plane in hcp.

Additionally, when considering the prismatic planes with the smallest Miller indices, $\{2\Bar{1}\Bar{1}0\}$ and $\{10\Bar{1}0\}$ can be seen to have a much lower PCSD in the \cref{fig:csl} b). However, the selection of GB planes cannot be entirely described by PCSD. In several cubic materials, no dependence of GB plane selection on PCSD is seen \cite{Rohrer2004}. In brass and nickel, a systematic GB plane analysis demonstrated that the dependence of interplanar spacing of the GB planes, d$_{eff}$, is the more critical criterion than PCSD in the selection of the GB planes \cite{KIM2005633}. For a symmetric GB, the d$_{eff}$ is same as the d-spacing of the GB plane, while for an asymmetric GB, the d$_{eff}$ is given by:
\begin{equation}
    d_{eff} = \frac{(d_{1}\: + \:d_{2})}{2}
    \label{eq:d_eff}
\end{equation}
where $d_1$ and $d_2$ are the d-spacing of the GB planes for each grain. The d$_{eff}$ is a means to generalize the d-spacing criterion for the asymmetric boundaries \cite{Randle2005}. The energy of an unrelaxed boundary increases as the d-spacing decreases, because the atoms with the shortest d-spacing contribute the most to the boundary energy. Consequently, low-index GB planes due to their larger d-spacing are preferred \cite{Randle1999}. In terms of probability, there are many possible asymmetric GBs and limited symmetric GBs. But once a low-index plane for one grain is fixed, the index of the other plane is constrained by the misorientation. As a result, most asymmetric GBs have a low d$_{eff}$ and are hence unfavorable. This is why symmetric GB planes appear more frequently than random distribution. In case of prismatic planes of Ti, $\{10\Bar{1}0\}$ has the largest d-spacing (2.55~\AA) followed by $\{2\Bar{1}\Bar{1}0\}$ (1.475~\AA). This leads to the largest d$_{eff}$ for the asymmetric $(10\bar{1}0)$ // $(11\bar{2}0)$ GB to be 2.01~\AA, which is much higher than 0.409~\AA\ for $\{\bar{7}520\}$ plane. In fact, another asymmetric GB $(\bar{3}210)$ // $(\bar{4}5\bar{1}0)$ is only 3$^{\circ}$ in-plane rotation away from the symmetric $\{\bar{7}520\}$. The d$_{eff}$ for $(\bar{3}210)$ // $(\bar{4}5\bar{1}0)$ is 0.76~\AA, that is also larger than the 0.409~\AA\ for $\{\bar{7}520\}$. In-spite of that, the symmetric GB plane, $\{\bar{7}520\}$ is observed more frequently in the present study and also in Mg and Ti-64 alloy \cite{OSTAPOVETS2014102, KELLY201622}. Other than the influence of extra-ordinarily high PCSD, one can anticipate that the atomic structure of the GB and the interaction energy of individual facets has a vital influence in the GB plane selection. It is now well established that GBs act as a 'phase' in themselves and have an atomic structure that is distinct from the two abutting grains \cite{Meiners2020, Frolov2013}. In all of existing literature that discusses the rules for selection of the GB \cite{Randle1999, Rohrer2004, KIM2005633}, the atomic structure of GB has never been taken into consideration. However, the structure of the GB can play a major role in determining its thermodynamic properties. The structure of the GB would also determine its ability to accommodate defects, equilibrium solute segregation and their influence on the GB energy. Therefore, although a combination of PCSD and d$_{eff}$ can be used to argue the stability of certain GB planes, further investigation of the atomic structure of the $\{\bar{7}520\}$ is needed to establish the reason for stability of the high-index symmetric GB plane over the low-index asymmetric plane, which is an ongoing work.

\subsection{Grain boundary segregation}

Solute segregation at GBs is known to largely influence the mechanical behaviour of metals either negatively, such as by embrittlement, or positively, by pinning the GBs and thus restricting grain growth, thereby strengthening it \cite{lejcek2010grain, Hofmann1996, Kirchheim2007}. Solute segregation can also lead to faceting of the GB \cite{DONALD1979, Peter2021}. 
The segregation of Fe in Ti is a known phenomenon. Random high angle GBs in Ti have been shown to be stabilized by the segregation of Fe in the substitutional sites \cite{Aksyonov2017}. Fe segregation has also been reported in commercial Ti alloys with no analysis of the influence of GB type on the segregation behaviour \cite{Ding2020, bermingham2009, OUTLAW1994143}. In the present study, Fe was observed to segregate preferentially in few selected symmetric facets in $\Sigma$13 (0001) Ti GB, as seen in \cref{fig:Fe_seg_eds}. All the symmetric facets were apriori expected to show isotropic behaviour as they belong to the same family of planes and are bound by the same two grains. The observed peculiar anisotropy is to a first approximation rationalized using the PCSD. The GB planes $(\bar{7}520)$ and $(\bar{5}\bar{2}70)$ have a higher PCSD and are seen to have lean or no segregation whereas, the GB planes $(\bar{5}7\bar{2}0)$, $(\bar{7}250)$, and $(\bar{2}\bar{5}70)$ have a lower PCSD and have much higher Fe segregation. Clearly, a lower PCSD seems to favour Fe solute segregation in Ti. %This is the first report of anisotropic segregation in any GB having the same misorientation ($\Sigma$ value) and the same GB plane. To add to the confirmation of above hypothesis of the dependence of faceting and segregation on PCSD, the asymmetric $(10\bar{1}0)$ // $(11\bar{2}0)$ GB revealed a continuous segregation of Fe all along the GB length (Supp. fig. 1).

Beyond characterizing the GB plane, it is of a great interest to quantify the amount of segregation. Such studies began in early 1980s using surface analysis techniques like Auger electron spectroscopy (AES) and secondary ion mass spectroscopy (SIMS) of the fracture surfaces \cite{Suzuki1981, LEJCEK2003, FUKUSHIMA1982753, Walther2004}. With manifold advancement of the analytical power of STEM, EDS and electron energy loss spectroscopy (EELS) are now regularly used to quantify lean segregation of even less than 0.01 monolayer at the GB \cite{doig1982, MULLER19961637, KEAST19993999, MULLER19961637}. More recently, state-of-the-art STEM techniques and atom probe tomography have been utilized in a correlative fashion to obtain spatial and chemical information from the same region down to almost atomic scale \cite{Christian2018, Zhou2016a, raabe2014grain, BABINSKY201495, stoffers2015, HERBIG201898}. A similar approach was used here to site-specifically lift-out a symmetric GB and quantify the Fe segregation as seen in \cref{fig:APT}. Although, the EDS clearly shows Fe segregation and partitioning to different GB segments, it needs however long counting times and thus may cause redistribution of Fe like seen in other metallic systems \cite{peter_liebscher_kirchlechner_dehm_2017}. Therefore, APT is used for the quantification of solute concentration. Additionally, the distribution of concentration in the third-dimension cannot be obtained in the STEM-EDS. It also helps to find any additional scarce impurity present in the material. From the atom distribution in \cref{fig:APT}, a uniform distribution of Ti and O is confirmed. Although the concentration of O is high, $\sim$28at.\%, it is measured to be uniformly distributed over the entire tip volume. The high O content in the film is due to the high solubility limit (32 at.\%) of O in Ti \cite{Murray1987}. Because the deposition chamber had a vacuum of only 2.2 $\times$ 10$^{-6}$~mbar, and Ti is widely used as a getter for O, the film is expected to have a high dissolved O content. However, no oxides or other secondary phases are detected. Most importantly, the distribution of O is not altered at the GB and can therefore be assumed to not have an influence on the Fe segregation.
Although the GB plane cannot be found from the APT data, we know that the film has only $\Sigma$13 GBs, with only two possible GB planes. As discussed earlier, the symmetric \{${\Bar{7}250}$\} GB is present $\sim$20 times more often than the $(10\bar{1}0)$ / $(11\bar{2}0)$ asymmetric variant, therefore we assume the captured GB to be the symmetric \{${\Bar{7}250}$\} GB. This assumption is also emphasised by the fact that the area density distribution in the \cref{fig:APT} h) shows the Fe distribution is restricted to $\sim$40-50 nm, which is about the same as the length of symmetric facet, as seen in \cref{fig:Facet}. Also, the areal density distinctly shows that Fe is uniformly distributed across the facet / GB plane. The decrease in the density of the Fe in the right part of the GB can be attributed to the commencement of the adjacent facet which is devoid of Fe. A striking similarity confirming the Fe-segregation restricted to symmetric GB segment can be seen in both \cref{fig:Fe_seg_eds} d), e) and \cref{fig:APT} e), f).
The subsequent measurement of interfacial excess using a 'ladder diagram' gave a $\Gamma_{Fe}$ of 1.25 at/nm$^2$ or 0.2 monolayer. The GB$_{excess}$ measurement has been used as a method to scrutinize phase formation in the GBs \cite{Maugis2016}. 0.2 monolayer corresponds to one in every five atoms at the GB plane being Fe. %Assuming that all Fe atoms are restricted to the GB plane, the stoichiometry corresponds to a Ti$_{0.8}$Fe$_{0.2}$ intermetallic BCC phase \cite{Sumiyama1986}.
With the spatial resolution of APT, there is no way to ascertain if the present observation is GB segregation (either by strain or chemically induced by bonding/charge transfer) or a GB phase transformation. Based on the local arrangement and bonding of Fe, its influence on the material properties could largely vary. This necessitates the requirement of additional experiments using STEM to discern the atomic structure of the GB with and without Fe, which is an ongoing work. %Moreover, the measurement of GB excess properties for all other frequently observed GBs will pave way for GB phase diagrams and GB engineering in the future.

\section{Conclusion}

Following the deposition of bicrystalline Ti thin films, faceting and Fe-segregation in $\Sigma13$ [0001] GBs of Ti thin films are explored here for the first time. Following are the important findings from this work:
\begin{enumerate}
\item A novel template based thin film deposition pathway for obtaining columnar grains containing tilt GBs of Ti is established using pulsed magnetron sputtering on SrTiO$_3$ (001) substrates at 600$^\circ$C followed by post-annealing at the same temperature for 2~h, 4~h and 8~h.
\item The ion current density during magnetron sputtering is seen to modify the texture completely from mixture of $(10\bar{1}1)$ and $(0002)$ to only $(0002)$ owing to the high adatom surface diffusion.
\item EBSD analysis reveals a bicrystalline film with mesoscopically curved GBs having a very high fraction of $\Sigma13$ [0001] CSL orientation. Such a textured film with well defined CSL GBs is demonstrated for the first time. 
\item The seemingly maze-like tilt GBs are verified to be columnar in nature and are shown to be faceted frequently into symmetric \{$\bar{7}520$\} and sporadically asymmetric $\{10\Bar{1}0\}$ // $\{2\bar{1}\bar{1}0\}$ facets. The selection of GB planes during faceting is considered to be a combination of high planar coincidence site density and high effective interplanar spacing (d$_{eff}$).
\item EDS analysis reveals a distinct preferential Fe segregation in every alternate \{$\bar{7}520$\} GB facet while \{$57\bar{2}0$\} GB remains Fe-lean. This is explained by the difference in the planar coincidence site density of the two planes within the same plane family which is unique to the basal plane six-fold symmetry in the hcp structure.
\item The Fe content is measured using atom probe tomography. Fe-segregation is observed to be limited to a fraction of the GB area.
\item The anisotropic segregation is expected to influence second phase nucleation, GB migration, and other critical material properties opening scope of similar investigations in other GBs and other hcp metals. %The influence of Fe segregation on GB atomic structure and GB migration is a part of the follow up study.

\end{enumerate}

\section*{Acknowledgement}
VD and CHL acknowledge the funding from KSB Stifftung. GD acknowledges funding from the European Research Council (Grant no. 787446-GB-CORRELATE). The authors would like to thank Dr. Baptiste Gault for his assistance with the APT experiments.

\bibliographystyle{ActaMatnew-2}
\bibliography{Paper2_citations}

\end{document}